\documentclass[useAMS,usenatbib]{mnras}
\usepackage[utf8]{inputenc}
\usepackage{amsmath}
\usepackage{amsfonts}
\usepackage{amssymb}
\usepackage{graphicx}
\usepackage{epstopdf}
\usepackage{listings}
\usepackage{multirow}

\usepackage{color}
\usepackage{xcolor}

\title{Impact of the external radiation field on the structure and evolution of supernova remnants}

\author[M.~Romero et al.]
{
M.~Romero$^{1}$, Y.~Ascasibar$^{1}$, J.~Palou\v{s}$^{2}$, R.~W\"{u}nsch$^{2}$, and M.~Moll\'{a}$^{3}$ \\
$^{1}$ Departamento de F\'{i}sica Te\'{o}rica, Universidad Aut\'{o}noma de Madrid, E-28049 Madrid, Spain\\
$^{2}$ Astronomical Institute, Czech Academy of Sciences, Bo\v{c}n\'{i} II 1401/1, 141 00 Praha 4, Czech Republic\\
$^{3}$ Departamento de Investigaci\'{o}n B\'{a}sica, CIEMAT, Avda. Complutense 40. E-28040 Madrid, Spain
}

\pagerange{\pageref{firstpage}--\pageref{lastpage}}
\pubyear{2021}

\newcommand{\pderiv}[2]{\frac{\partial #1}{\partial #2}}
\newcommand{\e}{\epsilon}

\newcommand{\Nabla}{\vec{\nabla}}
\newcommand{\Mach}{\mathcal{M}}

\newcommand{\cloudyCommand}[1]{\textsc{#1}}

\newcommand{\externalRadiationField}{ERF}

\begin{document}

\maketitle
\label{firstpage}

\begin{abstract}
We carry out 1D hydrodynamical simulations of the evolution of a spherically symmetric supernova remnant (SNR) subject to an external radiation field (\externalRadiationField) that influences the cooling and heating rates of the gas.
We consider homogeneous media with ambient hydrogen number densities $n_{\rm H,0}$ of $0.1$ and $1$ cm$^{-3}$ permeated by an average radiation field including the cosmic microwave, extragalactic, and Galactic backgrounds, attenuated by an effective column density $N_{\rm H,eff}$ from $10^{18}$ to $10^{21}$~cm$^{-2}$.
Our results may be classified into two broad categories:
at low $N_{\rm H,eff}$, the \externalRadiationField\ presents little absorption in the ultraviolet (ionising) regime, and all the 'unshielded' cases feature an equilibrium temperature $T_{eq} \sim 7000$~K below which the ambient gas cannot cool further.
In this scenario, the SNR develops a nearly isothermal shock profile whose shell becomes thicker over time.
At higher $N_{\rm H,eff}$, the \externalRadiationField\ is heavily absorbed in the UV range, yielding a roughly constant heating function for temperatures $\lesssim 10^4$ K.
These `shielded' cases develop a thin, cold and dense shell throughout their evolution.
Energy and momentum injection to the medium do not change significantly between both scenarios, albeit luminosity is higher and more uniformly distributed over the shell for unshielded SNR.
\end{abstract}


\section{Introduction}
\label{sec_Introduction}

Supernovae (SNe) are one of the key ingredients of galaxy formation and evolution~\citep{NaabOstriker17} due to the injection of energy, momentum and chemical elements to the surrounding medium.
They are one of the main physical {\sl feedback} mechanisms on the subsequent star formation processes occurring in galaxies, which has been an open problem since the earliest studies in the field.

The effect of a supernova explosion depends on the physical state of the interstellar medium (ISM).
In general terms, each SN ejects an energy of about 10$^{51}$\,erg, and the supernova remnant (SNR) undergoes several evolutionary phases during its lifetime.
The initial one is known as free expansion or ejecta-dominated, where the ejected mass is higher than the swept-up ISM mass.
A reverse shock inside the SNR that thermalizes the interior~\citep[][]{TrueloveMcKee99, FerreiraJager08, TangChevalier17}, yielding to the so-called Sedov-Taylor phase~\citep{Taylor50, Sedov59}.
The total energy of the SNR is still approximately constant, and analytical solutions for both the evolution of the shock and the radial profiles of the thermodynamic quantities are very well known \citep[e.g.][]{OstrikerMcKee88}.

As radiative losses become important, energy drops significantly and momentum injection eventually becomes roughly constant in time.
Early studies of SNR focused on 1D spherically symmetric numerical simulations~\citep{Chevalier74, OstrikerMcKee88, Cioffi+88, Franco+94, Thornton+98, Shelton98}.
More recent studies about this subject focus on less idealized conditions~\citep[][and references therein]{KimOstriker15a, Miao+15, Sarkar+20}
or in presenting semi-analytical models that capture the main physical properties observed in the simulations~\citep[e.g.][]{Haid+16, LeahyWilliams17, DiesingCaprioli18, Jimenez+19}.

The internal structure of the SNR becomes more complex during the radiative phase, as a cold, dense shell develops just behind the forward shock, which is prone to both physical and numerical instabilities~\citep{Blondin+98, Badjin+16}.
During the shell formation, the total luminosity emitted by the SNR reaches its maximum value for its whole evolution~\citep[e.g.][]{Chevalier74, Thornton+98, Jimenez+19}, and the shell plays a dominant role in driving the SNR dynamics.
The precise internal structure of the shell is also extremely important in order to estimate the spectral signatures of the SNR, and it has been extensively studied in the literature~\citep[e.g][]{Chevalier+80, Terlevich+92, DopitaSutherland96, Allen+08}.

Radiative heating and cooling are thus extremely important to fully understand the evolution of SNR as well as to predict their observable properties.
In particular, the treatment of cooling at low temperatures (i.e. $T<10^4$~K) can alter the shell structure and shock dynamics ~\citep{Badjin+16}.
Many previous works are based on a predefined cooling function, often based on collisional ionisation equilibrium~\citep[e.g.][]{SutherlandDopita93, Grackle}.
For heating, they either cut cooling below some threshold temperature around $\sim 10^4$\,K~\citep[e.g.][]{Cioffi+88, Thornton+98, Martizzi+15, Slavin+15, Pittard19} or use a constant value of $2\times 10^{-26}$\,erg\,s$^{-1}$~\citep[][]{WalchNaab15, KimOstriker15a, Haid+16} based on equation (5) from ~\citet{KoyamaInutsuka02} \citep[which, in turn, fits the results obtained in][]{KoyamaInutsuka00}.

More recently, several studies have considered the effects of radiative transport on the evolution of SNR at various levels of detail, showing that radiation from the progenitor star~\citep{WalchNaab15, Geen+15} or self-radiation of the SNR gas~\citep{Sarkar+20} have a minor effect on the dynamics of the SNR and the total amount of energy and momentum injected into the ambient medium, although they affect the heating, cooling, and chemical composition of the gas inside the remnant.

Here we consider the external radiation field (\externalRadiationField) that pervades the ISM, originating from both the overall stellar population of the Galaxy, as well as the extragalactic radiation background.
This radiation field is known to be a critical ingredient to determine the physical conditions of the ISM \citep[e.g.][]{Wolfire+95}, and we would like to assess its impact on the physical properties of the SNR and their evolution.

In order to address this question, we perform one-dimensional hydrodynamical simulations of supernova explosions in a homogeneous medium.
We propose an approximate methodology to account for radiative transport, assuming that the simulated region is permeated by a uniform average radiation field.
This may be used to compute the cooling and heating functions to be applied into the numerical simulations.
The computation of the average intensity of the radiation field is described in Section~\ref{sec_Environment}.
The setup of our numerical simulations is discussed in Section~\ref{sec_Simulations}, and the results are presented in Section~\ref{sec_Evolution}.
Section~\ref{sec_Discussion} discusses our approximation, comparing the results with previous approaches commonly used in the literature, and focusing on potential caveats and limitations.
Our main conclusions are briefly summarised in Section~\ref{sec_Conclusions}.

\section{Approximate treatment of the ISM}
\label{sec_Environment}

In order to model the interstellar gas as accurately as possible, we must consider the main heating and cooling processes.
In this work, we use the photoionisation code \cloudyCommand{Cloudy} \citep{Cloudy17} to pre-compute the corresponding heating and cooling tables that can be later used by any hydrodinamical simulation code.
A schematic summary of our method, including examples of the input scripts used in the present work, is provided in Appendix~\ref{app_StepByStep}

\subsection{Physical conditions of the gas}
\label{ssec_Physical conditions}

We consider two different values of the hydrogen number density in the SNR environment, $n_{\rm H,0}=0.1$ and $1$~cm$^{-3}$.
In order to approximate radiative transfer, we consider that the radiation field from the ionising stars has been attenuated by an effective hydrogen column density $N_{\rm H,eff}$ before reaching the gas particles within the region where the supernova explodes.
This parameter reflects the effective optical depth towards the unabsorbed radiation sources, and hence it may depart from the simple expectation $N_{\rm H,eff} \sim n_{\rm H,0} L$, where $L$ is the spatial extent of the simulation box.
For instance, it would be larger if most of the surrounding ionising stars are heavily obscured, while it would be lower if they find a clear line of sight towards the ambient medium.
Rather than modelling the complex processes and geometrical details of radiative transfer, we simply
select the values $N_{\rm H,eff}=\{10^{18},\ 10^{19},\ 10^{20},\ 10^{21}\}$~cm$^{-2}$, as representative of typical optical depths for the warm atomic and ionized medium of the Milky Way \citep{Wolfire+95, Ferriere98,Ferriere01}.

The chemical composition of the gas is specified by the \cloudyCommand{abundances ISM} built-in keyword~\citep[mainly based on tables 3 and 5 of][respectively]{CowieSongalia86,SavageSembach96}.
This command implicitly adds \cloudyCommand{grains ISM} to consider the presence of dust grains.
In this work, we do not investigate the injection of additional dust particles by the supernova ejecta nor the destruction of the existing ISM grains by the SNR blast wave~\citep[for an interested reader, see e.g.][ and references therein]{Slavin+15,MartinezGonzalez+20,Priestley+21}.
Instead, we consider an additional set of runs adding the \cloudyCommand{no grains} keyword after the \cloudyCommand{abundances ISM} command to bracket the maximum possible impact of dust heating and cooling (or lack thereof) on the evolution of the SNR.

Finally, we also consider the presence of cosmic rays by adding the \cloudyCommand{cosmic rays background} command, which adopts an $H^0$ cosmic ray ionization rate of $2 \times 10^{-16}~$s$^{-1}$ and $4.6\times 10^{-16}$~s$^{-1}$ for $H_2$ secondary ionization rate~\citep[see][]{GlassgoldLanger74, Indriolo+07}.

\subsection{Average radiation field}
\label{ssec_AverageRadiationField}

To model the effects of the \externalRadiationField, we add the command \cloudyCommand{table ISM} to represent the unattenuated radiation field emitted by the stellar population~\citep{Black87}, supplemented by \cloudyCommand{Table HM12} to include also the extragalactic background~\citep{HaardtMadau12} and \cloudyCommand{CMB} for the cosmic microwave background, in which a blackbody radiation field in strict thermodynamic equilibrium at $T = 2.725$~K is assumed.
The last two commands are evaluated at redshift~$0$.

\begin{figure}
    \hspace{-0.5 cm}
    \centering
    \includegraphics[width=0.5\textwidth]{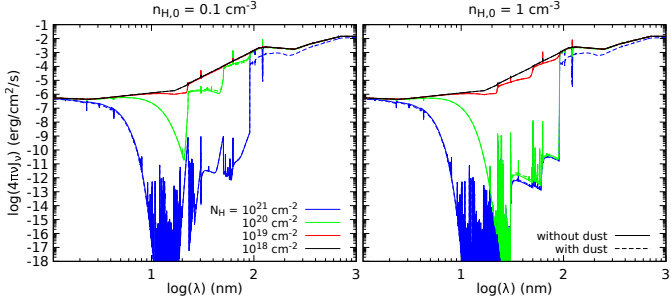}
    \caption{Transmitted continuum spectra given by \cloudyCommand{Cloudy} for each run. Each colour represent a different column density, while dashed and continuous lines represent if there is dust or not, respectively.}
    \label{Fig_cutSpectra}
\end{figure}

\begin{figure}
    \hspace{-0.5 cm}
    \centering
    \includegraphics[width=0.5\textwidth]{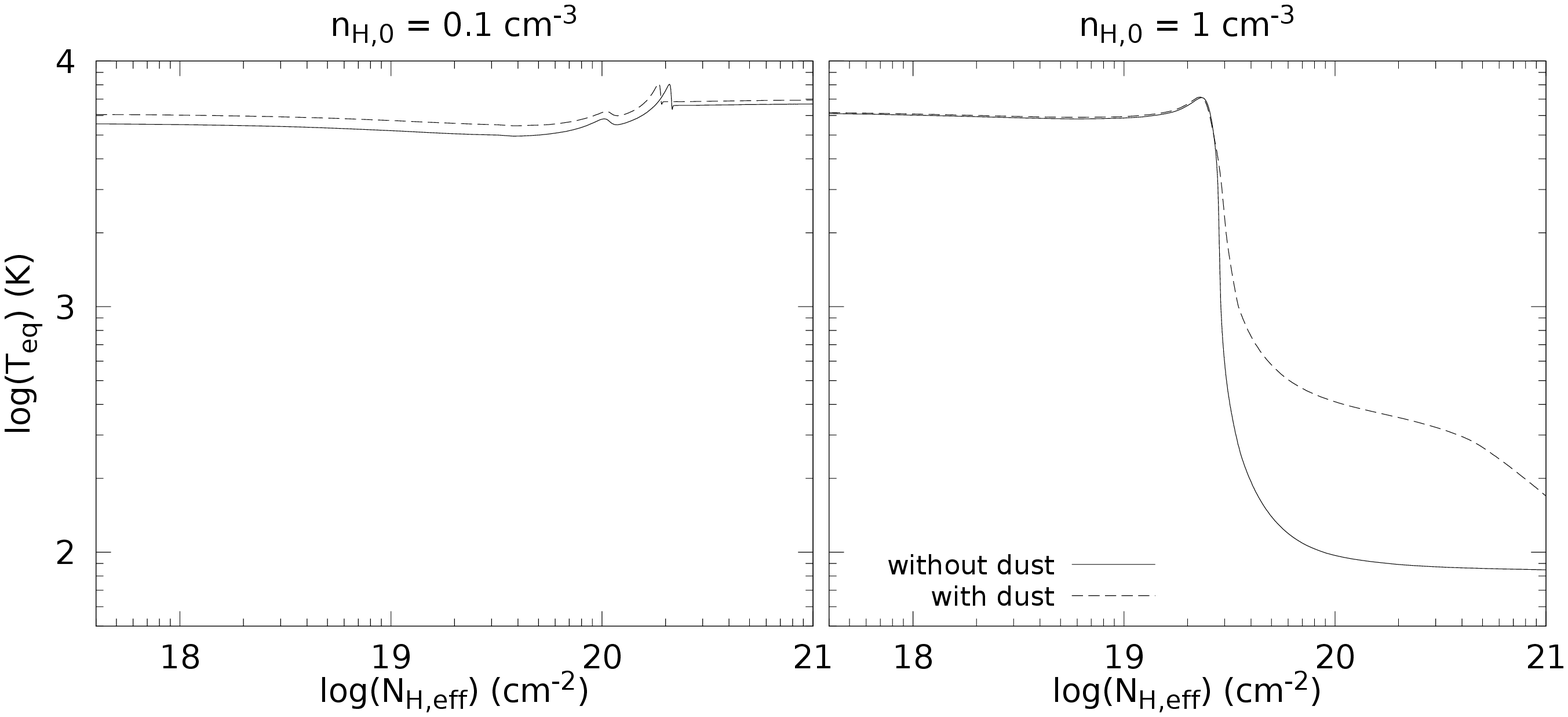}
    \caption{Equilibrium temperature of the cloud as a function of effective column density.}
    \label{Fig_Teq}
\end{figure}

The main output that we obtain from \cloudyCommand{Cloudy} is the transmitted continuum spectra (Figure~\ref{Fig_cutSpectra}) and the equilibrium temperature $T_{eq}$ when the desired $N_{\rm H,eff}$ is reached (Figure~\ref{Fig_Teq}).
Although the spectrum becomes gradually more absorbed as the column density increases, the overall effect on the final radiation field can be qualitatively understood by classifying the environmental conditions into two groups:
\begin{enumerate}
    \item In simulations with $n_{\rm H,0} = 1$~cm$^{-3}$ and $N_{\rm H,eff}=10^{20}$ or $10^{21}$~cm$^{-2}$, as well as $n_{\rm H,0} = 0.1$~cm$^{-3}$ and $N_{\rm H,eff}=10^{21}$~cm$^{-2}$, most radiation in the UV range (hydrogen-ionising photons) is absorbed before reaching the environment of the SNR.
    We will refer to this group as the \emph{shielded} case.
    \item The remaining simulations do show much less absorption in the UV spectral range compared with the previous ones, and therefore they will be referred to as the \emph{unshielded} case.
\end{enumerate}
In general, the equilibrium temperature of the unshielded cases is of the order of $\sim 7000$~K, whereas it drops by more than an order of magnitude for the shielded cases.
This is not true for $n_{\rm H,0} = 0.1 cm^{-3}$ and $N_{\rm H,eff}=10^{21}~cm^{-2}$, that keeps a warm $T_{eq}$ due to the effect of the cosmic-ray heating, but the cooling and heating functions, on the other hand, are similar to the other shielded cases.

The presence of dust in the ISM has only a minor effect on the final spectrum in the vicinity of the SNR, which is almost fully determined by the effective column density $N_{\rm H,eff}$ and the ambient density $n_{\rm H,0}$, but it will have a significant impact on the cooling and heating functions at high gas temperatures.

\section{Hydrodynamical simulations}
\label{sec_Simulations}

\subsection{Code overview}
\label{ssec_codeOverview}

We wrote an Eulerian finite-volume 1D hydrodynamical code with Adaptive Mesh Refinement (AMR), heavily based on~\citet{DoumlerKnebe10} cosmological magnetohydrodynamic code \citep[see also][]{Ziegler04,Ziegler05}.
The whole set of equations, which are solved in spherical coordinates with the method two of~\citet{WangJohnsen17}, for mass, $\rho$, total energy, $\e$, and momentum, $\rho v$, densities are:
\begin{eqnarray}
& \pderiv{\rho}{t} + \frac{1}{r^{2}}\pderiv{}{r}(r^{2} \rho v) = 0
\label{eq_Euler_mass} \\
& \pderiv{\e}{t} + \frac{1}{r^{2}}\pderiv{}{r}[r^{2} v(P+\e) ] = \dot{u}_H -\dot{u}_C
\label{eq_Euler_energy} \\
& \pderiv{}{t}(\rho v) + \frac{1}{r^{2}}\pderiv{}{r}[r^{2} (\rho v^2 + P) ] = 0
\label{eq_Euler_momentum}
\end{eqnarray}
where $\dot{u}_H$ and $\dot{u}_C$ denotes energy densities changes due to heating and cooling, respectively (see~\ref{ssec_CoolingHeating}).
Pressure is defined as:
\begin{equation}
P = (\gamma - 1)[\e - \frac{1}{2}\rho v^2],
\label{eq_Pressure_energy}
\end{equation}
To avoid a negative pressure due to truncation errors, which are a common issue for highly supersonic fluids (such as thin-shell formation in SNR), we adopt the \citet{Ryu+93} method by adding an additional equation to the system of~\eqref{eq_Euler_mass}-\eqref{eq_Euler_momentum} with a source term accounting energy loses/gains:
\begin{equation}
\pderiv{S}{t} + \frac{1}{r^{2}}\pderiv{}{r}(r^{2}Sv) = \frac{(\gamma - 1)}{\rho^{\gamma-1}}(\dot{u}_H - \dot{u}_C)
\label{eq_Euler_entropy}
\end{equation}
where $S$ is known as modified entropy.
\begin{equation}
S \equiv \frac{P}{\rho^{\gamma - 1}}
\label{eq_Entropy}
\end{equation}
We solve equations~\eqref{eq_Euler_energy} and~\eqref{eq_Euler_entropy} simultaneously, and we check, for each cell, if the fraction of thermal energy is lower than a parameter $\delta$ and the cell is \emph{not} undergoing a shock
\begin{eqnarray}
& \frac{\e - \frac{1}{2}\rho v^2}{\e} < \delta
\label{eq_Thermal_fraction} \\
& \frac{1}{r^2}\pderiv{}{r}(r^{2} v) \ge 0 
\label{eq_NoShock_condition1}
\end{eqnarray}
If both conditions are true for one cell, we compute pressure with~\eqref{eq_Entropy} and also rewrite energy density of said cell as:
\begin{equation}
\e = \frac{1}{2}\rho v^2 + \frac{S \rho^{\gamma - 1}}{\gamma - 1}
\end{equation}
Otherwise, we rewrite $S$ instead using its definition~\eqref{eq_Entropy}.
A small $\delta$ (even as small as zero) will protect any code from giving negative pressure values, and the dynamics (e.g. Energy and momentum input into the ISM) can be tracked accurately.
However, we opted to use $\delta = 0.3$, which is, at least, an order of magnitude higher than the one used in other studies~\citep[][among others]{Ryu+93,DoumlerKnebe10,Gentry+17} to track temperature accurately (i.e. without oscillations due to truncation errors) as well.

\subsection{Initial conditions, AMR and timesteps}
\label{ssec_IC}

Each simulation consists of a sphere of radius $8\times 10^{20}~$cm ($\simeq 260\,$pc) with the equilibrium temperature found at the end of the cloud simulated by \cloudyCommand{Cloudy} in previous section, which matches with the desired column density (see Figure~\ref{Fig_Teq}).
We subdivide the simulation sphere in $52000$ cells, which is a resolution of roughly $\Delta r \simeq 0.005$~pc
\footnote{Higher resolutions and/or lower densities than $0.1~cm^{-3}$ are risky, as the resolution can become unphysical since the mean free path can be longer than the cell length \citep[see appendix of][]{Fierlinger+15}}.
To handle that, we also implement an AMR scheme with $10$ levels of refinement to manage the resolution and performance, taking previous values as the maximum resolution.
The criteria for refinement is based on energy density variations between cells. Thus, if
\begin{equation}
|\frac{\e_{i+1} - \e_{i-1} }{\e_i}| < 0.005,
\label{eq_Refine_criterion} 
\end{equation}
we merge cells $i$ and $i+1$ provided that both have the same resolution.
On the other hand, we split cell $i$ when the above expression gives a value higher than $0.1$. If that cell is already at maximum resolution, we split its closest neighbours unless they are already at the same resolution.

We start the simulation at maximum resolution modelling a Sedov-Taylor phase at $t=0$, thus skipping the ejecta-dominated phase that comes before.
To achieve that, we change energy density, inspired by the recipe given by~\citet{TrueloveMcKee99}, and modified entropy for the first $400$ cells closest to the center as:
\begin{eqnarray}
& \e_i = \e_0 + \frac{E_{0}}{\Delta V_i}  f(\frac{r_i}{5\Delta r_i})
\label{eq_Energy_injection} \\
& S_i = (\gamma - 1)\frac{e_i}{\rho_i^{\gamma - 1}}
\end{eqnarray}
where $E_{0} = 10^{51}\,$erg, $\Delta V_i$ is the cell volume in cm$^{3}$, $r_i$ and $\Delta r_i$ are the cell location and width, respectively, in cm; and $f(x) \propto e^{-x^2}$, normalized to $400$ cells.

Finally, we use as timestep the smallest, for all cells, of the dynamical and the cooling timesteps.
On one hand, dynamical timestep per cell is based on the known Courant condition: that the fluid should not travel farther than a given fraction $\varepsilon_{_{CFL}}$ of the size of the own cell $i$
\begin{equation}
\Delta t_i \le \varepsilon_{_{CFL}}\frac{\Delta r_i}{\max(v_i,c_{s,i})},
\label{eq_Courant_condition}
\end{equation}
where $c_{s,i}$ is the sound speed of the cell and $\varepsilon_{_{CFL}}$ is a parameter that should be less than $0.5$ to guarantee stability \citep{DoumlerKnebe10}.
On the other hand, we define a cooling time for each cell as
\begin{equation}
    t_{_C,i} = \frac{P_i}{(\gamma-1) \dot{u}_{C,i}}
\end{equation}
and define the cooling timestep as $\varepsilon_{C}\cdot t_{_C,i}$ where $\varepsilon_C$ have the same purpose as $\varepsilon_{_{CFL}}$ to guarantee stability.

\subsection{Cooling and heating rates}
\label{ssec_CoolingHeating}

\begin{figure}
    \centering
    \includegraphics[width=0.5\textwidth]{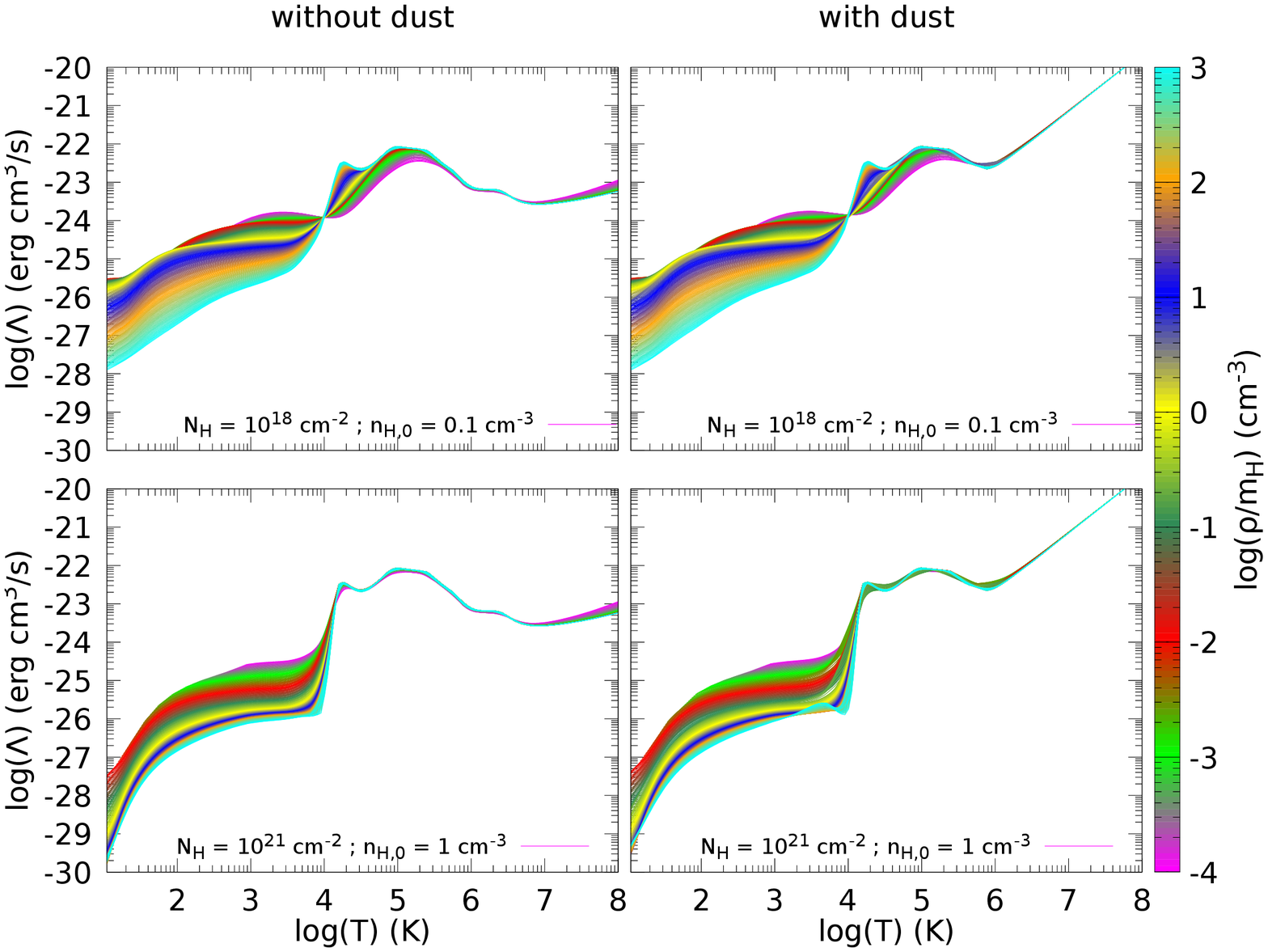}
    \includegraphics[width=0.5\textwidth]{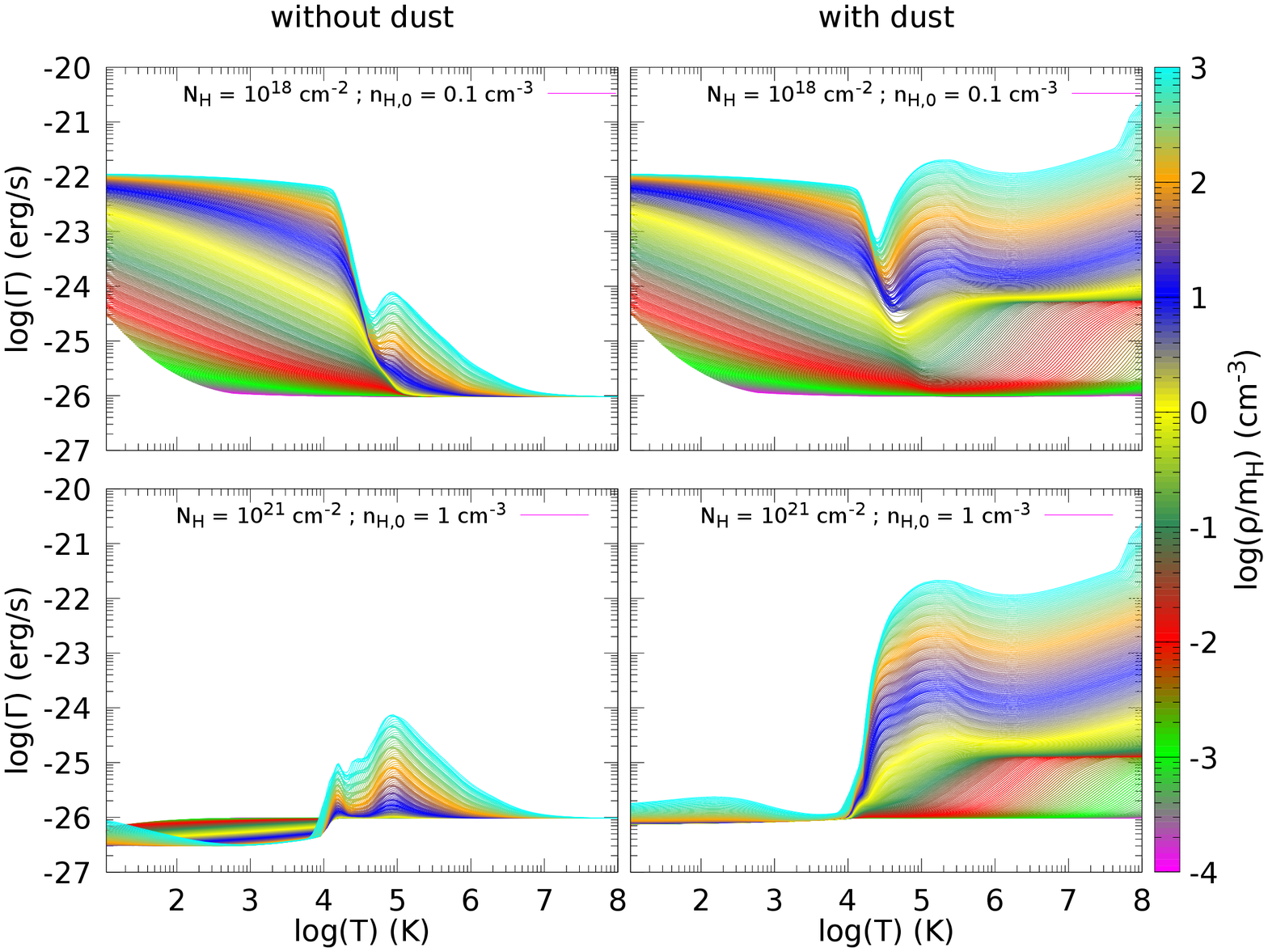}
    \includegraphics[width=0.5\textwidth]{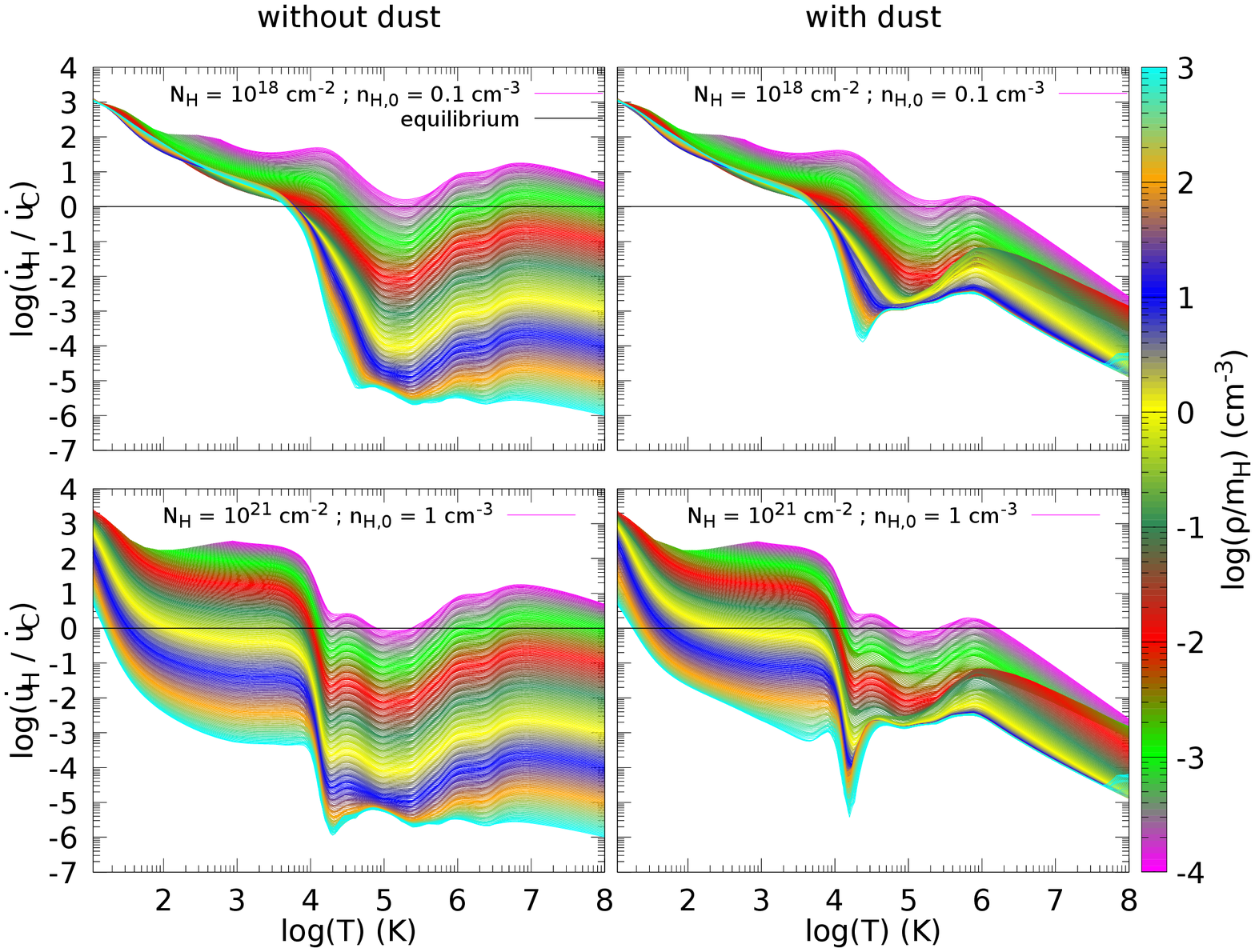}
    \caption{Cooling ($\Lambda \equiv (\frac{\mu m_H}{\rho})^2\dot{u}_C$, upper panel) and heating ($\Gamma \equiv \frac{\mu m_H}{\rho} \dot{u}_H$, middle panel) functions of four representative cases, as well as the ratio $\dot{u}_H/\dot{u}_C$ (bottom panel).
    Top and bottom rows within each panel show unshielded and shielded cases, respectively.}
    \label{Fig_CoolingFunction}
\end{figure}

In order to handle heating and cooling, we assume that each simulation cell is illuminated by the mean radiation field computed in section~\ref{ssec_AverageRadiationField} for the selected ambient hydrogen density and equilibrium temperature.
To do that, we perform a second \cloudyCommand{Cloudy} using the spectra of Figure~\ref{Fig_cutSpectra} as incident radiation field of the cloud, while keeping the other characteristics of the ISM (i.e. \cloudyCommand{cosmic rays background}, \cloudyCommand{abundances ISM} commands, and \cloudyCommand{no grains}, if applicable) identical to those of the ambient medium.
The relevant output, $\dot{u}_H$ and $\dot{u}_C$ at the illuminated face of the cloud, are saved as a function of the possible temperature (between $10$ and $10^9 K$) and hydrogen density (from $10^{-6}$ to $10^4~cm^{-3}$) that the gas cells may reach during the simulation (not to be confused with the initial values $T_{eq}$ and $n_{\rm H,0}$ of the ambient medium).

Then, we translate the output of this second \cloudyCommand{Cloudy} run,  $\dot{u}_C$ and $\dot{u}_H$ as a function of $n_H$ and $T$, to a set of three tables in terms of the total mass density $\rho$ and the pressure $P$ of the gas cell, which are the physical variables internally used by the hydrodynamical code.
At each simulation step, these tables, provided in Table~\ref{tab_TableSample} in Appendix~\ref{app_StepByStep} for the interested reader, are interpolated using bilinear interpolation of their logarithms to produce the $\dot{u}_H$ and $\dot{u}_C$ needed for equations~\eqref{eq_Euler_energy} and~\eqref{eq_Euler_entropy}.

We show in Figure~\ref{Fig_CoolingFunction} the cooling function, defined as $\Lambda \equiv (\frac{\mu m_H}{\rho})^2\dot{u}_C$ (where $\mu$ denotes the mean atomic weight, and $m_H$ the hydrogen mass), the heating function $\Gamma \equiv \frac{\mu m_H}{\rho} \dot{u}_H$, and the ratio between the heating and cooling rates $\dot{u}_H/\dot{u}_C$.
Top rows display two representative examples of the unshielded cases, while bottom rows show shielded ones.
Left columns show dust-free media, and the effect of dust particles is illustrated on the right columns.

In general terms, the presence of an \externalRadiationField\ increases both the cooling and heating rates at temperatures below $\sim 10^4$~K, whereas dust tends to increase them above $\sim 10^5$~K and $\sim 10^4$~K, respectively.

As can be seen in the lower panel, differences in the heating rate are irrelevant for $T > 10^4$~K and $\frac{\rho}{m_H} > 10^{-2}$~cm$^{-3}$, because heating is absolutely negligible compared to cooling in this regime.
At high temperatures, though dust becomes the main cooling agent, and it dramatically enhances the cooling rate at $T>10^6$~K.
A detailed treatment of the destruction of dust grains would be of the utmost importance in order to accurately model the cooling rate, but we consider that our two extreme cases may illustrate the associated uncertainties on the evolution of SNR.

In any case, the gas will eventually cool down towards the equilibrium temperature on a relatively short time scale\footnote{All our \cloudyCommand{Cloudy} simulations assume colissional ionisation equilibrium. Especially for hot diffuse gas, the recombination time may be longer than the cooling and/or dynamical time scales. Under those conditions, non-equilibrium effects should be taken into account \citep[e.g.][]{DopitaSutherland96, GnatSternberg07, Vasiliev13}.}.
For temperatures lower than $10^4$~K, there are obvious qualitative differences between the shielded and unshielded cases.
If the ISM radiation field is heavily attenuated, the cooling function is very close to collisional ionisation equilibrium, and the heating function is similar to the $2 \times 10^{-26}$~erg~s$^{-1}$ advocated by \citet{KoyamaInutsuka02}.
The ratio $\dot{u}_H/\dot{u}_C$ between the heating and cooling rates is very sensitive to the gas density, but it is very flat on the temperature range between $\sim 10^2$ and $10^4$~K.
For that reason, the equilibrium temperature of the ISM changes drastically between these extremes at densities around $\sim 1$~cm$^{-3}$.

This contrasts with unshielded cases, where heating is indisputably more important than cooling for $T < 10^4$~K, and therefore no gas can exist for long periods of time below that temperature.
This situation is akin to the classical models where the radiation field is ignored, but cooling is manually switched off below a certain threshold of the order of the equilibrium temperature.

\section{SNR Evolution}
\label{sec_Evolution}

\subsection{Integrated quantities}
\label{ssec_SNRcharact}

\begin{figure}
    \centering
    \includegraphics[width=0.5\textwidth]{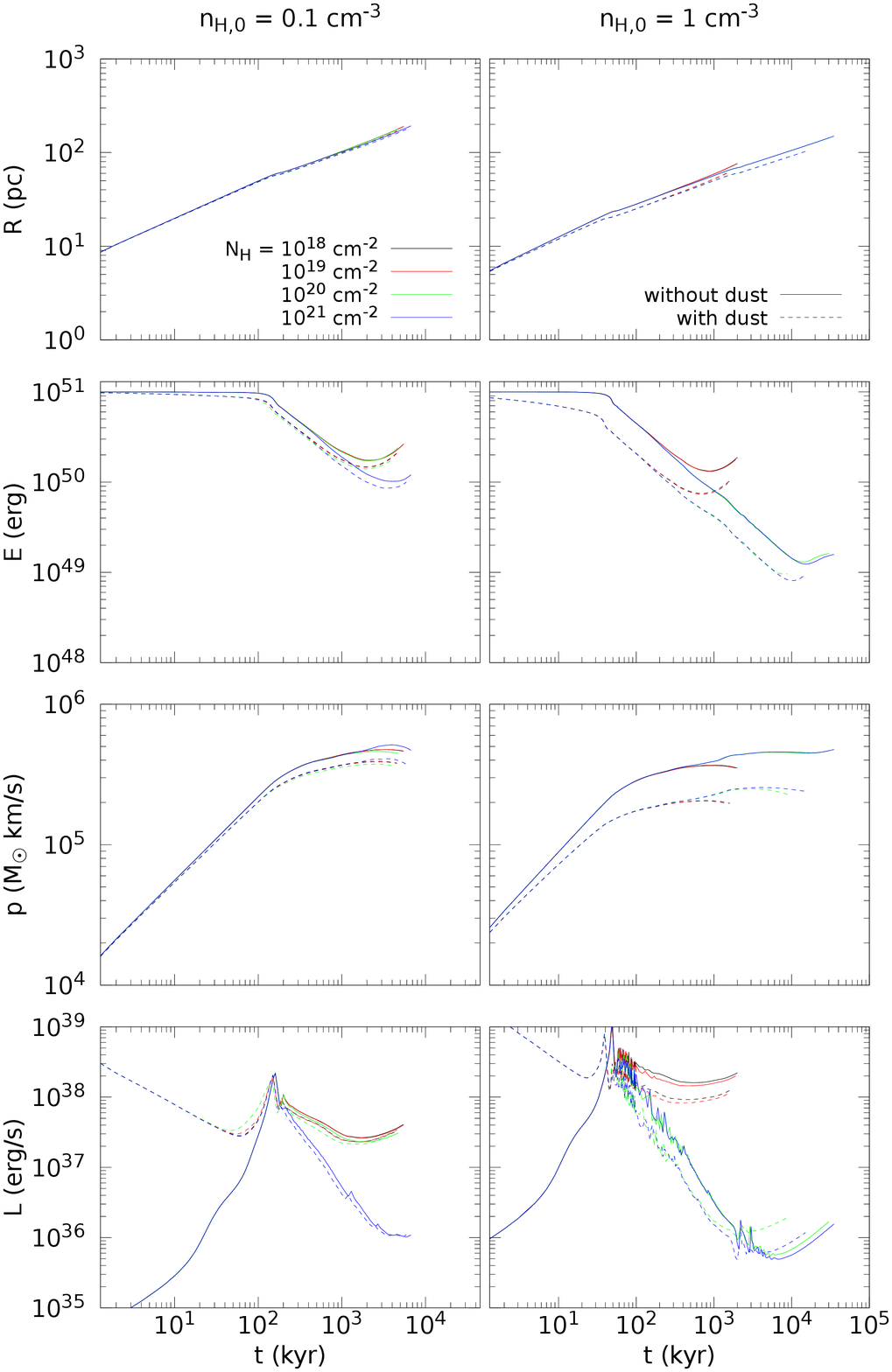}
    \caption{Temporal evolution of the shock radius, $R$, total energy, $E$, momentum, $p$, and luminosity, $L$, as function of time for all SNR cases.}
    \label{Fig_evolution}
\end{figure}

In order to characterize the evolution of the SNR, we first find the location of the shock radius $R(t)$, which we identify with the farthest distance from the initial explosion at $t=0$ that fulfills the following criteria:
\begin{eqnarray}
\Nabla \cdot \vec{v} &<& 0 
\label{eq_ShockCondition1} \\
\Nabla T \cdot \Nabla \rho &>& 0
\label{eq_ShockCondition3} \\
\mathcal{M} &>& 1.3
\label{eq_ShockCondition2}
\end{eqnarray}
The first condition implies that the cell is being compressed, whereas the second one discriminates a shock wave from a cold front discontinuity.
Ideally, condition~\eqref{eq_ShockCondition2} should be $\Mach > 1$, but we choose a higher value to avoid false positives caused by spurious numerical oscillations in the thermodynamical variable~\citep{SchaalSpringel15, Pfrommer+17}.
Once $R$ is found for each time, we compute total energy, momentum, and luminosity of the SNR from the volume integrals
\begin{eqnarray}
E(t) &=& \int_{0}^{R(t)} \e~4\pi r^2 dr
\label{eq_IntE_def} \\
p(t) &=& \int_{0}^{R(t)} (\rho v)~4\pi r^2 dr
\label{eq_pValue_def} \\
L(t) &=& \int_{0}^{R(t)} \dot{u}_C~4\pi r^2dr
\label{eq_L}
\end{eqnarray}

The evolution of all these quantities is plotted in Figure~\ref{Fig_evolution}.
From the point of view of SNR dynamics, there is barely any significant difference between shielded and unshielded cases as long as the shock remains strong ($\mathcal{M}\gg 1$).
For instance, at $1$ Myr, differences in radius and momentum are less than $5$ and $10$ percent, respectively.
Energy, on the other hand, show higher differences, up to a factor of $2$, because the shock is no longer strong in the unshielded case.
In the weak-shock regime, at the latest stages of the evolution, the equilibrium temperature and pressure, $T_{eq}$ and $P_{eq}$, of the ambient medium become relevant, explaining the changes in $R$, $E$ and $p$ at late times that one can observe in Figure~\ref{Fig_evolution}.
Basically, the sound speed is lower in the shielded cases, and therefore the weak shock regime, as well as the associated upturn in the total energy within the SNR, are delayed with respect to the unshielded cases.
These results suggest that a more complex heating and cooling scheme would not have a strong effect on the evolution of the SNR as a whole (i.e. shock radius, total energy and momentum), in agreement with the recent results reported by \citet{Sarkar+20}.

The evolution of the shielded cases is consistent with the results of previous studies based on the \citet{KoyamaInutsuka02} heating prescription \citep[e.g.][]{WalchNaab15, KimOstriker15a, Haid+16}, while unshielded cases match the evolution obtained by switching off cooling at a warm equilibrium temperature \citep[e.g.][]{Cioffi+88, Thornton+98, Martizzi+15, Slavin+15, Pittard19}.

However, the balance between heating and cooling processes is very different under both situations, even if the net effect is similar.
In the end, the SNR must dissipate the kinetic energy of the gas that is incorporated to the shock, $L_{\rm iso} = 2\pi\rho_0 R^2\dot R^3$ \citep[see e.g.][]{Cioffi+88}, which is roughly the same in both cases, given that the evolution of the shock radius is almost identical.
On the other hand, the energy absorbed from the radiation field is radically different for a shielded and unshielded environment, being significantly higher for the latter.
The heating rate is thus much higher, but the total cooling rate (i.e. the luminosity of the SNR) increases as well, until the bulk of the absorbed energy is immediately re-radiated by the gas, and it has a minor impact on the evolution of the shock radius and the integrated energy and momentum.

The presence or absence of dust particles does not alter this conclusion, although it plays a significant role in the SNR dynamics at early times, when the temperature of the hot interior is higher than $10^6$~K and dust is the major coolant.
If the supernova ejecta enriched the gas with newly synthesised dust and/or a fraction of the pre-existing grains survived the shock passage, radiative losses during this phase will not be negligible, at variance with the classical Sedov-Taylor regime.
Once the post-shock temperature drops below $\sim 10^5$~K, other cooling agents become dominant, and shell formation proceeds exactly as in the classical case.
The main difference is that the energy lost during early evolution is not negligible (about 10 percent for $n_{\rm H,0} = 0.1$~cm$^{-3}$ and 40 percent for $n_{\rm H,0} = 1$~cm$^{-3}$).
Shell formation happens slightly sooner in the dusty simulations, and early radiative losses result in a lower final energy and momentum injection into the ISM.

\subsection{Internal structure}

\begin{figure*}
    \centering
    \includegraphics[width=0.48\textwidth]{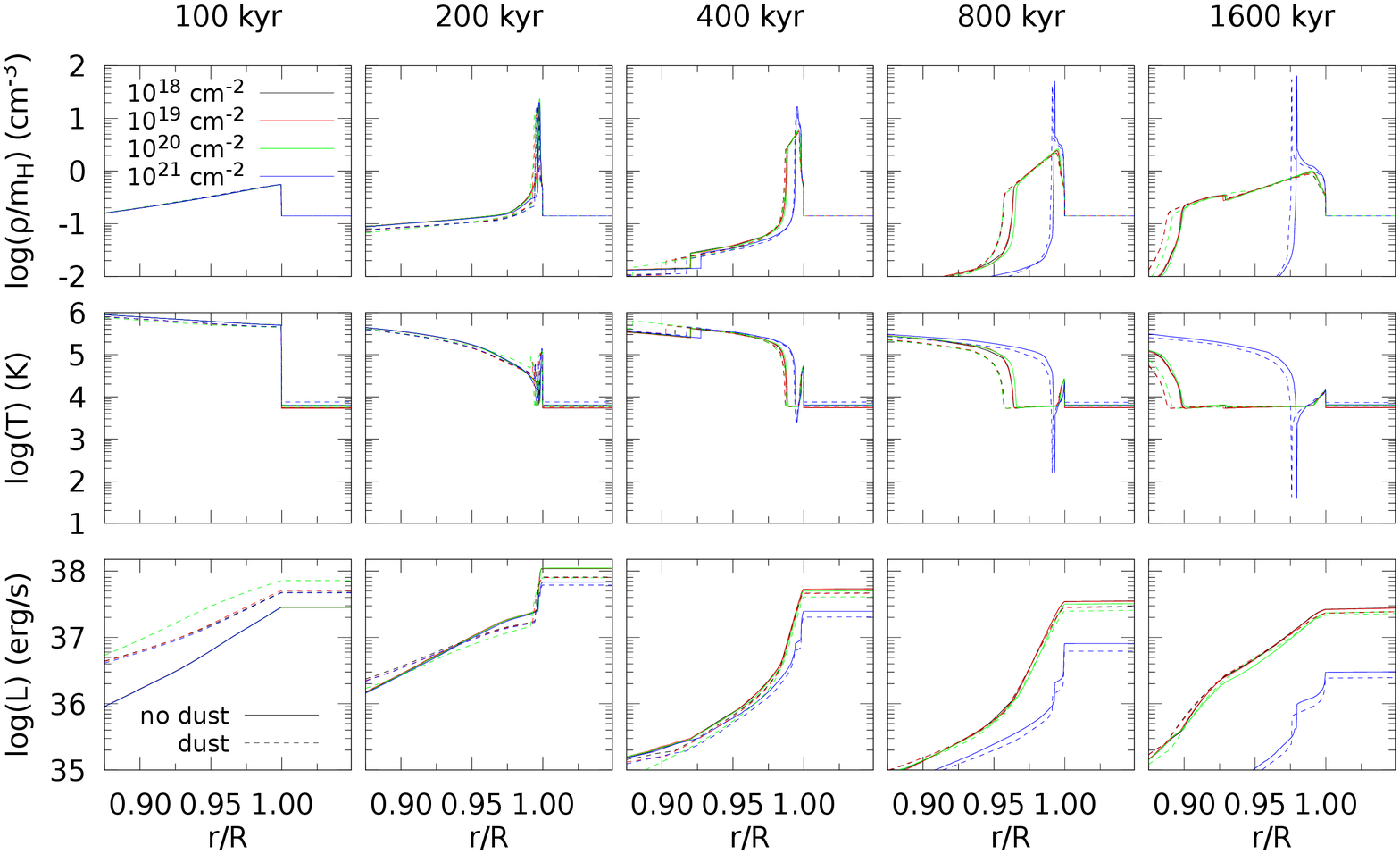}
    \includegraphics[width=0.48\textwidth]{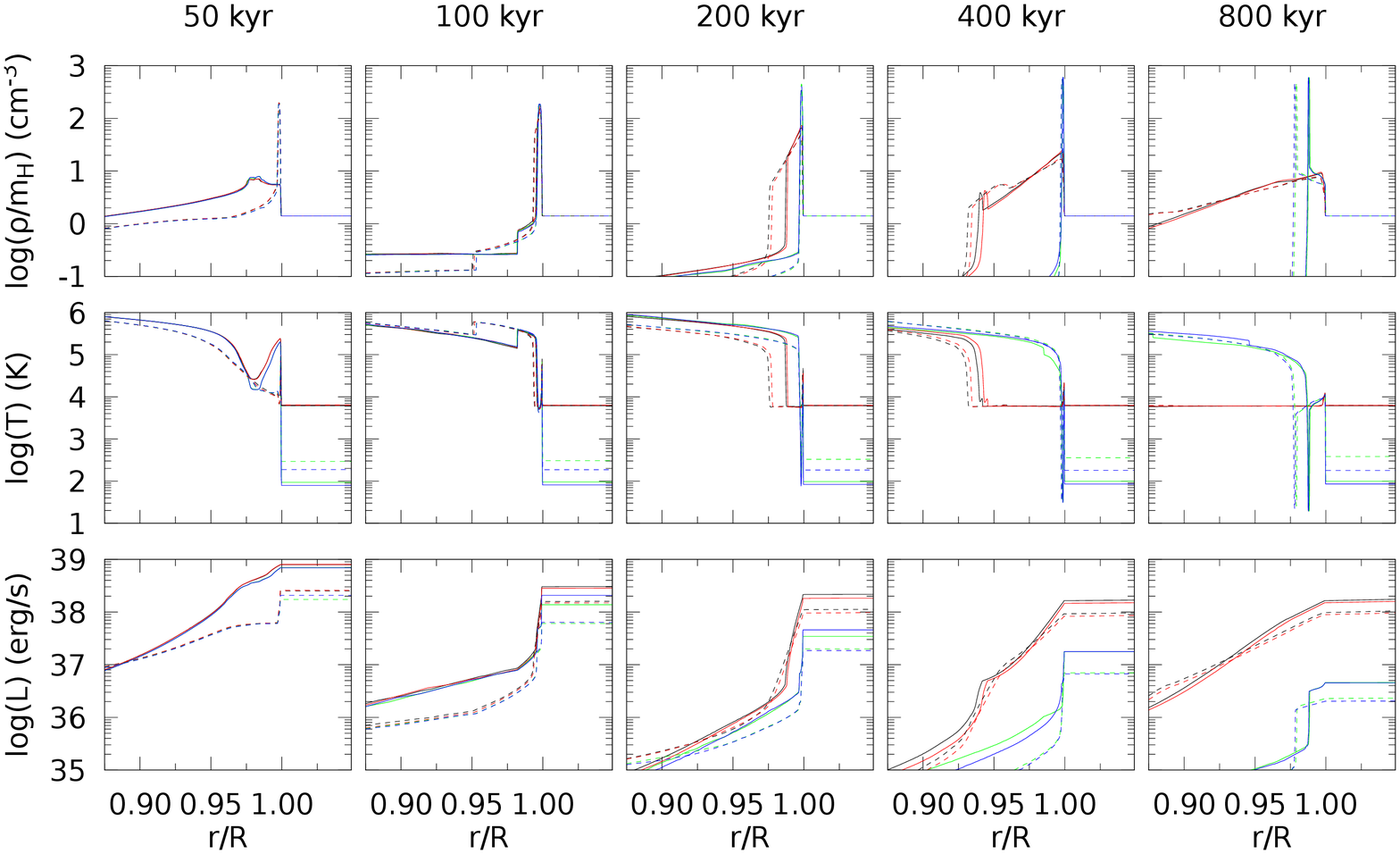}
    \caption{Integrated radial profiles of mass density, temperature, and luminosity normalized to the shock radius (i.e.: $1$ is $R$) for particular time outputs for all SNR simulations.
    Left panel are the results with initial hydrogen density of $n_{\rm H,0} = 0.1~cm^{-3}$, whereas right panel represents $n_{\rm H,0}=1~cm^{-3}$.
    Each color represent a column density.}
    \label{Fig_profiles}
\end{figure*}

Figure~\ref{Fig_profiles} shows the radial profiles of gas density, temperature and integrated luminosity near the shock as a function of $r/R$, where $r$ is the radial coordinate, for a few times between $50$ to $1600$ kyr, mostly focused on the radiative phase of the SNR evolution.

After the Sedov-Taylor phase, unshielded and shielded cases diverge significantly in the predicted structure of the post-shock shell.
Unshielded shells become thicker and less dense over time, displaying a structure that is similar to a classical isothermal shock.
On the other hand, shielded cases are better described by the infinitely thin-shell approximation over a long period of time.
Its density is much higher, and its temperature much lower, than the unshielded case.
At very late times, though, shielded cases eventually develop a warm region at $T\sim 10^4$~K after the shock, roughly similar (albeit much thinner) to the unshielded case.
The thin shell, about two orders of magnitude colder, is slightly displaced inwards, delineating the boundary between the warm region and the hot bubble.

As mentioned above, there are important differences in the total luminosity radiated by the shielded and unshielded cases after the shell is formed.
Moreover, the physical properties of the emitting gas and the structure of the emission are very different in both scenarios.
In the unshielded case, the emissivity is smoothly distributed over the shell, and cooling is dominated by the recombination and collisional emission lines characteristic of a $T\gtrsim 10^4$~K gas, above the equilibrium temperature.
In contrast, the (much lower) luminosity of the shielded case may be separated into the contribution of the gas cooling from the post-shock temperature to the equilibrium temperature, which is similar to the unshielded case, and the emission from the very dense, cold shell at $\sim 100$~K near the hot bubble.
The fractional contribution of the latter varies from with time, but it is always of the order of $\sim 20-50$ per cent.
The warm region between them at $T \sim T_{eq}$ has no significant contribution.

The physical reason between these differences can be understood from the ratio between the heating and cooling rates depicted in Figure~\ref{Fig_CoolingFunction}.
In the unshielded cases, is barely impossible for the gas to cool down below the equilibrium temperature, whereas a much wider range of stable configurations exist in the shielded scenario, where sufficiently dense gas is allowed to reach much lower temperatures.

Although dust particles may effect the evolution in time of the shock radius, they barely have any impact on the morphology of the shell during the strongly radiative phase.
Simulations including dust feature a shorter cooling time (see Figure~\ref{Fig_evolution}), and therefore the profiles shown in Figure~\ref{Fig_profiles} correspond to a more evolved state.
In addition, the presence of dust also affects the equilibrium temperature of the ambient medium (see Figure~\ref{Fig_Teq}), as well as the heating and cooling rates of the post-shock gas.
Although this plays a major role on the total luminosity radiated by the SNR, it has a minimal effect on its internal structure.

\section{Discussion}
\label{sec_Discussion}

\subsection{Astrophysical context}

Hydrogen number densities of the ambient medium $n_{\rm H,0}$ in the range $0.1$ to $1$ cm$^{-3}$ are fairly typical of the warm ionized and atomic medium, as well as the cold atomic medium in the Solar vicinity~\citep[e.g.][]{Ferriere01}.
Column densities towards the ionising sources $N_{\rm H,eff}$ between $3\times 10^{18}$ to $3\times 10^{20}$ cm$^{-2}$ are also representative of the local ISM, according to~\citet{Wolfire+95}.
We explore slightly higher column densities, up to $10^{21}$ cm$^{-2}$, in order to investigate possible trends within the shielded cases.
Although this value is arguably rather high for the Solar neighborhood (except for extreme cases, such as shielding by a nearby dense molecular cloud), it may occur in other environments.
The total hydrogen column density of the Milky Way disk may exceed $10^{21}$ cm$^{-2}$ for Galactocentric radii of $\sim 5$ kpc~\citep[see Figure 1 of][]{Ferriere01}, and observational data of~\citet{Leroy+08} show that other galaxies can reach values that are higher by more than an order of magnitude.
Although the total column density of the disk provides an upper limit to the absorbing column, we do think that $10^{21}$~cm$^{-2}$ is not entirely unrealistic.

Regarding the progenitor of the SNR, our set up is valid for an isolated explosion.
This case may correspond to a white dwarf exploding as SN Ia or a runaway OB star~\citep{GiesBolton89c}.
These events can take place anywhere within the ISM, albeit they are more likely to occur in the disk~\citep{JohnsonMacLeod63, Hakobyan+17}.
Typical core collapse SNe would explode in denser environments, but the presence of other stars and previous SNe would yield a much more complex scenario, whose details have recently been explored in the literature~\citep[see e.g.][]{KimOstriker15a,Vasiliev+17,Gentry+17,Gentry+18}.

\subsection{Implications}

The main result of the present work is that the most relevant effect of an \externalRadiationField\ is to set the heating and cooling functions, and thus the conditions of the ambient medium where the SNR propagates.
Broadly speaking, our numerical experiments can be classified in two different scenarios, that we have dubbed shielded and unshielded cases, separated by a hydrogen column density around $N_{\rm H,eff} \sim 10^{20}$ cm$^{-2}$, albeit there is some dependence with the ISM number density (see Figures~\ref{Fig_cutSpectra} and~\ref{Fig_Teq}).
We find that the evolution of the main global properties of the SNR is fairly robust and does not depend strongly on the details of the radiation field, with the only exception of the total luminosity, which may vary by more than an order of magnitude.
Moreover, our results also show that the internal structure of the shell is very different in the shielded and unshielded cases.

Observationally, these differences arise at very late times, of the order of $\sim 0.3-3$~Myr after explosion, where detecting SNR is challenging.
Surveys based on X-rays~\citep[e.g.][]{Long+10, Leonidaki+10, Sasaki+12} are better suited to trace the earliest stages of SNR evolution, whereas the latter phases of the radiative shells are more readily observed in the optical~\citep[e.g.][]{LeeLee14a, LeeLee14b} or radio~\citep[e.g.][]{Green14, DubnerGiacani15} bands.
Very old SNR have interacted so much with the inhomogeneous environment that only fragments of the shell may be observed.
There are some individual objects reported as probable candidates for very old SNR, such as e.g. G55.0+0.3~\citep{Matthews+98}, G106.3+2.7~\citep{PineaultJoncas00}, FVW172.8+1.5~\citep{Kang+12} or GSH 90-28-17~\citep{XiaoZhu14}.
All of them display average radii between 50 and 100~pc, consistent with our results, and show obvious signs of fragmentation.
With the exception of G106.3+2.7, they retain an approximately circular or elliptical shape on large scales.

Our results suggest that the \externalRadiationField\ has an important role in setting the overall luminosity of the remnants, especially in the optical and infrared regime.
Besides the temperature structure of the shell, the ionisation and population balance of the gas is also set by radiative equilibrium (hence the effect on the cooling function), and therefore not only the luminosity but also the optical and infrared~\citep[e.g.][]{Reach+06} emission line ratios will be different in the unshielded and shielded shells.

Furthermore, we think that the distinction between the shielded and unshielded scenarios may have implications regarding supernova feedback, even if the energy and momentum injection into the ISM are not significantly affected.
Specifically, the interaction of SNR with molecular clouds and star formation has been extensively studied under different conditions and strategies~\citep[e.g.][and references therein]{IffrigHennebelle15, Korolev+15, Kortgen+16, Lucas+20, Lu+20} and it is still an open problem.
In general terms, shocks propagating through a low-density medium (or channel) will be able to propagate far away from the original place of the explosion, whereas dense media will be more resilient to shock passage.
Albeit these results are robust with respect to an external radiation field, the physical state of the post-shocked gas is radically different in the warm ($T \sim 10^4$~K) thick shell of an unshielded SNR and its cold ($T \sim 100$~K) dense counterpart.

\subsection{Comparison with previous schemes}

To first order, our results support the validity of previous works that include an approximate treatment of the external radiation field through different modifications of the cooling and heating functions.
On the one hand, unshielded SNR are compatible with the prescription of switching off cooling below a certain temperature threshold.
Based on our findings, we would advocate to set that threshold to the equilibrium temperature of the ISM, which may be computed by a photoionisation code such as \cloudyCommand{Cloudy} from the desired \externalRadiationField, neglecting attenuation.
This approximation will be appropriate for an optically thin medium.
On the other hand, shielded SNR are consistent with a fixed value of the heating rate, although the precise numbers we find are slightly below the fiducial $\Gamma = 2\times 10^{-26}$ erg/s suggested by \citet{KoyamaInutsuka02}.
In order to estimate the appropriate value for the desired conditions, we propose to carry out a \cloudyCommand{Cloudy} simulation with a representative column density (the \cloudyCommand{extinguish} command should be enough for this purpose) that is sufficiently thick to place the gas in the shielded case.
The resulting heating rate will depend of the dust content, as well on any other heating sources (e.g. turbulence, cosmic rays) in addition to the radiation field, which are indeed dominant in the shielded scenario.

Although the results will be qualitatively similar, the approach used in the present work will be more accurate than the classical approximations, in the sense that the cooling and heating functions would be more realistic for the whole range of temperatures and densities reached in the simulation.
In particular, the heating rate will include the absorbed energy, as well as the effect of the \externalRadiationField\ on the population equilibrium of the gas.

\subsection{Caveats and potential improvements}

Nevertheless, there are important shortcomings and limitations to the proposed approach.
First, it should be noted that these calculations were made under the assumption of ionization equilibrium.
Non-equilibrium effects reduce the cooling function, from a factor of $2$ up to an order of magnitude, for temperatures lower than $10^6$~K~\citep{GnatSternberg07,Vasiliev13}, although the magnitude of these differences may be overestimated in presence of an extragalactic background~\citep{OppenheimerSchaye13}.
Furthermore, the recent work by~\citet{Sarkar+20} studied this particular issue and found that, at shell formation, when the temperature drops below $10^6$~K, the cooling function increases compared with those in ionization equilibrium (see their Figure 5).
These differences were reported to have a negligible effect into the dynamics and energetics of the SNR, although they cannot be neglected for predicting observables, such as emission spectra.

Even more importantly, instabilities that distort the shape of shell cannot be followed by a one-dimensional simulation.
According to the analytical calculation of~\citet{VishniacRyu89}, verified by the numerical results of~\citet{Blondin+98}, a realistic isothermal shock would be unstable at any Mach number higher than $\sim 3$ if the post-shock gas cannot cool below the ambient temperature, as in our unshielded conditions.
On the other hand, \citet{Pittard+05b} found that this limit drops to lower Mach numbers if the gas is allowed to cool further, as in our shielded case.
The distinction between both regimes is thus important in this particular context.

Furthermore, the interstellar medium is anything but homogeneous, which also contributes to break the assumption of spherical symmetry.
Nevertheless, it has been shown that an inhomogeneous medium does not alter considerably the net energy and momentum injection into the ISM, which is always within a factor of $\sim 2$ of the results obtained in the homogeneous, spherically-symmetric case~\citep[see e.g.][]{WalchNaab15, Martizzi+15,KimOstriker15a,Pittard19}.

If individual sources of radiation (i.e. stars close to the SN) were to be included, on-the-fly radiative transfer would be necessary.
Even in the spherically symmetric case, the assumption of an average radiation field would not hold anymore, as the radiation field would vary as a function of both position and time.
In addition, nearby stars\footnote{For illustration purposes, the local stellar density in the Solar neighbourhood is $0.1$ pc$^{-3}$~\citep{HolmbergFlynn01}.} would also enter the hot interior of the SNR as it increases in size, and their (almost unattenuated) radiation would reach the inner interface of the shell, potentially altering its morphology in the shielded scenario, which would cease to be valid.

On the other hand, unshielded SNR are more robust with respect to the details of the individual sources of radiation, as the whole medium is assumed to be optically thin, but they are prone to self-shielding of the shell, which can potentially become opaque.
According to Figure~\ref{Fig_Teq}, this transition occurs around $N_{\rm H,thick} \sim 2 \times 10^{20}$ cm$^{-2}$ for $n_{\rm H,0} = 0.1$~cm$^{-3}$ and $N_{\rm H,thick} \sim 3\times 10^{19}$ cm$^{-2}$ for $n_{\rm H,0} = 1$ cm$^{-3}$.
We can make a rough estimation of the associated SNR radius $R_{\rm thick}$ at the time of the transition by assuming that all the swept-up mass is concentrated inside an infinitely thin shell, $N_{\rm H,thick} \sim \frac{n_{\rm H,0}}{3} R_{\rm thick}$.
This yields $R_{\rm thick} \sim 2$~kpc for $n_{\rm H,0} = 0.1$~cm$^{-3}$ and $R_{\rm thick} \sim 30$~pc for $n_{\rm H,0} = 1$~cm$^{-3}$.
Thus, full radiative transfer, including absorption within the shell, should be taken into account when modelling a SNR with an unshielded ERF propagating in a dense medium.

One may extend the procedure presented in section~\ref{sec_Environment} to less idealized environments by taking into account that the incident radiation field will vary across the simulation box along the evolution of the SNR.
Rather than fully solving radiative transport, it should be possible to devise a fast algorithm that uses the column density towards individual sources to identify localized shielded and unshielded regions, neglecting the narrow transition (see Figure~\ref{Fig_Teq}) between the two cases.
Average radiation fields, as well as cooling and heating functions, could then be estimated within each region from previously computed tables.

\section{Conclusions}
\label{sec_Conclusions}

This work considers the role of the external radiation field (\externalRadiationField), from both Galactic and extragalactic sources, on the physical properties of SNR.
More precisely, we simulate the evolution of a spherically-symmetric explosion in a uniform medium, where the \externalRadiationField\ is approximated by an average radiation field that determines the gas heating and cooling rates as a function of density and temperature.
We use the photoionisation code \cloudyCommand{Cloudy}~\citep{Cloudy17} to estimate the average radiation field that reaches the region of interest after traversing a certain column density of intervening gas and to compute the corresponding cooling and heating tables.
We select two different values of the ambient density, $n_{\rm H,0} = 0.1$ and $1$~cm$^{-3}$, and four different column densities $N_{\rm H,eff} = \{10^{18},10^{19},10^{20},10^{21}\}$~cm$^{-2}$.
We performed two sets of simulations, with and without dust, to bracket the extreme cases where all/no dust grains survive the SNR shock.
Our main conclusions can be briefly summarised as follows:
\begin{enumerate}
    \item The effects of the \externalRadiationField\ can be classified into two different scenarios:
    unshielded cases ($N_{\rm H,eff} \lesssim 10^{20}$ cm$^{-2}$ for $n_{\rm H,0} = 0.1$ cm$^{-3}$ and $N_{\rm H,eff} \lesssim 10^{19}$ cm$^{-2}$ for $1$~cm$^{-3}$), where the ionising radiation prevents the gas from cooling below an equilibrium temperature around $\sim 7000$ K, and shielded cases ($N_{\rm H,eff} \gtrsim 10^{21}$ cm$^{-2}$ for $n_{\rm H,0} = 0.1$ and $N_{\rm H,eff} \gtrsim 10^{20}$ cm$^{-2}$ for $1$~cm$^{-3}$) where the UV continuum is heavily absorbed.
    The latter yield an approximately constant heating function $\Gamma$ for $T<10^4$ K, and dense gas can reach temperatures below $100$~K.
    \item The \externalRadiationField~does not change significantly the energy and momentum input into the environment, in agreement with previous results in the literature.
    \item In contrast, radial profiles become different at late times: unshielded cases develop a warm shell that becomes thicker and fades away over time, while shielded cases develop a much denser, colder shell at the interface with the inner hot bubble.
    \item Due to these differences, as well as the heating contributed by the absorbed photons, the emission of the SNR at different wavelengths is very different in both scenarios.
    \item The presence of dust does not have a significant impact on the above classification nor the internal structure of the SNR. However, it changes the heating and cooling functions for $T>10^5$ K, and radiative losses before shell formation lower the energy and momentum injection into the ISM (up to a factor of $\sim 2$ for $n_{\rm H,0} = 1$~cm$^{-3}$).
\end{enumerate}

To sum up, we consider that the proposed approximation based on an average radiation field provides an efficient way to account for the main effects of the \externalRadiationField.
Based on our results, we conclude that this is an important ingredient in order to determine the observable properties of old SNR, such as luminosities and emission line ratios.
The physical state of the gas within the shell may also have profound implications regarding supernova feedback on subsequent star formation.

\section*{Acknowledgments}

We are indebted to Alexander Knebe for providing useful suggestions during the development of the code.
We also like to thank 
Iker Millan and Lluis Galbany for helpful discussions.
This work has been supported by the Spanish Ministry of Economy and Competitiveness (MINECO) through the MINECO-FEDER grants AYA2016-79724-C4-1-P, AYA2016-79724-C4-3-P, PID2019-107408GB-C41 and PID2019-107408GB-C42.
JP and RW acknowledge the support by the Czech Science Foundation Grant 19-15480S and by the project RVO:67985815.

\section*{Data availability}

Tables of cooling and heating rates obtained from \cloudyCommand{Cloudy} are available as online supplementary material. Other data underlying this article will be shared on reasonable request to the corresponding author.


\bibliographystyle{mnras}
\bibliography{references}

\appendix

\section{Step by step process to generate cooling and heating tables}
\label{app_StepByStep}

\begin{table*}
\hspace*{-1cm}
    \begin{tabular}{ c l l l}
    \hline
    & Average radiation field & Cooling and heating tables & SNR simulations \\ 
    \hline \hline
    Inputs & Ambient hydrogen number density, $n_{\rm H,0}$. & Range of possible number densities $n_H$. & Ambient mass density, $\rho_0$. \\
    & Chemical composition. & Chemical composition. & Ambient temperature, $T_{eq}$. \\
    & Unattenuated radiation field $J_{\nu,0}(\lambda)$. & Average radiation field, $J_\nu(\lambda)$. & Cooling and heating tables.\\
    & Column density to ionising field, $N_{\rm H,eff}$. \\
    \hline
    Methods & \cloudyCommand{Cloudy}. & \cloudyCommand{Cloudy}. & 1D hydrodynamical code. \\
    \hline
    Outputs & Transmitted continuum, $J_\nu(\lambda)$. & Cooling and heating tables. & SNR radial profiles. \\
    & Equilibrium temperature, $T_{eq}$. & Total mass density and pressure. & SNR integrated magnitudes. \\
    \hline
    Explained in & Section~\ref{sec_Environment}. & Section~\ref{ssec_CoolingHeating}. & Sections~\ref{ssec_codeOverview} and~\ref{ssec_IC}. \\
    \hline \hline
    \end{tabular}
    \caption{Outline of the steps taken in this work. Starting with from left (Average radiation field) to right columns.}
    \label{tab_summary}
\end{table*}

In this appendix we explain, from a technical point of view, how to compute the cooling and heating tables needed in the hydrodynamical simulations. This scheme is summarized in Table~\ref{tab_summary}.

First, we write a \cloudyCommand{Cloudy} input script that specifies the desired hydrogen number density $n_{\rm H,0}$ of the ambient medium, as well as its chemical composition.
Destruction of dust grains by the SNR shock can be modelled by adding \cloudyCommand{no grains} to the \cloudyCommand{abundances ISM} command.
The unattenuated ionising radiation field, $J_{\nu,0}(\lambda)$, is represented by a combination of built-in tables, corresponding to the CMB, extragalactic background and ISM radiation fields.
The purpose of this step is to compute the average radiation field that reaches the region of interest after crossing a typical column density, $N_{\rm H,eff}$.
In addition, cosmic-ray ionisation with the default rate \citep{GlassgoldLanger74,Indriolo+07} is also included.
An example input script is shown below:
\begin{verbatim}
title radiation field example
#
# INPUT PARAMETERS:
# Ambient density and chemical composition
hden -1 # log(n_H0/cm^-3)
abundances ISM no grains
# Unattenuated ionising field J_nu0
table ISM
table HM12 redshift 0 # extragalactic background
CMB
# Cosmic-ray ionisation
cosmic rays background
# Total hydrogen column density
stop column density 21 # log(N_{\rm H,eff}/cm^-2)
# Control parameters
stop temperature off
iterate to convergence
#
# OUTPUT:
save transmitted continuum "Field" last # J_nu
save abundances "Species" last # number densities
save overview "Overview" last # n, T, X_ion, etc.
\end{verbatim}
For the different range of parameters, the only changes to this example are the numeric values in the \cloudyCommand{hden} and \cloudyCommand{stop column density} commands, as well as the presence/absence of \cloudyCommand{no grains} in the \cloudyCommand{abundances} command.
We specify \cloudyCommand{stop temperature off} to override the default stopping condition and allow the computation to proceed into the $T<10^4$~K regime.
We also added the \cloudyCommand{iterate to convergence} to achieve optimal accuracy in the output.
On output, the code saves the transmitted continuum intensity $J_\nu(\lambda)$ that we need for the next step, as well as the main physical properties of the gas.
For our purposes, we are interested in the number densities of the different chemical elements (from the `Species' file), as well as the equilibrium temperature of the ambient medium, which can be extracted from the `Overview' file.
The keyword `last' in the output commands is introduced in order to save only the last iteration, where convergence has been achieved.

Next, we write a set of \cloudyCommand{Cloudy} inputs for a range of number densities between $10^{-6}$ to $10^4$ cm$^{-3}$, all of them using the transmitted continuum $J_\nu(\lambda)$ (i.e. 'Field' file) obtained above to represent the average radiation field reaching a gas cell.
An example input script of this step would be:
\begin{verbatim}
title cooling table example
#
# INPUT PARAMETERS:
# Gas density and chemical composition of gas cell
hden 0.4  # log(n_H/cm^-3)
abundances ISM no grains
# Average field J_nu
table read "Field" scale 1
# Cosmic-ray ionisation
cosmic rays background
# Control parameters
set nmaps 100  # number of temperatures
#
# OUTPUT:
save map "Cooling", zone 0 range 1 to 9
\end{verbatim}
This script computes the heating and cooling rates for a gas cell with $n_H \simeq 2.5$~cm$^{-3}$ and the same chemical composition as the ambient medium.
These rates (as well as the number density of free electrons and other useful quantities) are saved in the \cloudyCommand{`Cooling'} file in units of erg/cm$^3$/s for a range of temperatures between $10$ and $10^9$ K, subdivided in $100$ logarithmic steps.

Finally, we process the \cloudyCommand{Cloudy} output to compute the total number density, mean atomic weight, pressure and mass density using data from the 'Cooling' and 'Species' files:
\begin{eqnarray}
 n &=& n_e + \sum_{Z=1}^{36} n_Z \\
 \rho &=& \sum_{Z=1}^{36} A_Z n_Z m_H \\
 \mu &=& \frac{\rho}{n m_H} \\
 P &=& n k_B T
\end{eqnarray}
where $Z$, $A$, $m_H$ and $k_B$ are the atomic and mass numbers, hydrogen mass and Boltzmann constant, respectively.
Bear in mind that $\rho$ is not a function of temperature and can be used as parameter in the same way as $n_H$.

We provide in Table~\ref{tab_TableSample} the cooling and heating rates, total and electron number densities, mean atomic weight, pressure and adiabatic index, $\gamma$, for the example above.
The complete version, as well as the tables obtained for the other environments of this work are given in electronic format.

\begin{table*}
\begin{tabular}{ c c c c c c c c c}
    \hline
    $n_H$ (cm$^{-3}$) & $T$ (K) & $\dot{u}_C$ (erg/s/cm$^3$) & $\dot{u}_H$ (erg/s/cm$^3$) & $\mu$ & $n$ (cm$^{-3}$) & $n_e$ (cm$^{-3}$) & $P$ (erg/cm$^3$) & $\gamma$ \\
    \hline \hline
    $1\times 10^{-6}$ & $10.0$ & $7.3267\times 10^{-37}$ & $1.7109\times 10^{-32}$ & $0.79874$ & $1.7579\times 10^{-6}$ & $6.5939\times 10^{-6}$ & $2.4271\times 10^{-21}$ & $1.6667$ \\
    $1\times 10^{-6}$ & $12.023$ & $7.1232\times 10^{-37}$ & $1.7276\times 10^{-32}$ & $0.79097$ & $1.7752\times 10^{-6}$ & $6.7667\times 10^{-6}$ & $2.9468\times 10^{-21}$ & $1.6667$ \\
    $1\times 10^{-6}$ & $14.456$ & $6.9542\times 10^{-37}$ & $1.7433\times 10^{-32}$ & $0.78374$ & $1.7916\times 10^{-6}$ & $6.9304\times 10^{-6}$ & $3.5758\times 10^{-21}$ & $1.6667$ \\
    \vdots & \vdots & \vdots & \vdots & \vdots & \vdots & \vdots & \vdots & \vdots \\
    \hline \hline
\end{tabular}
\caption{Sample of the cooling and heating table for the case $n_{\rm H,0} = 0.1$ cm$^{-3}$ and $N_{\rm H,eff} = 10^{21}$ cm$^{-2}$ without dust. The full version of this table, as well as the tables for the other cases shown in this work, are given in electronic format.}
\label{tab_TableSample}
\end{table*}

\label{lastpage}
\end{document}